\documentclass[sigconf]{acmart}
\def\plaintitle{Privacy Issues of Behavioral User Profiling via Mouse Tracking}
\def\catchytitle{My Mouse, My Rules}
\def\plainauthor{Luis A. Leiva, Ioannis Arapakis, Costas Iordanou}
\def\plainkeywords{Mouse Cursor Tracking; User Profiling; Privacy; Ethics}

\usepackage[utf8]{inputenc}
\usepackage{microtype}
\usepackage{enumitem}
\usepackage{xcolor}
\usepackage{bm}
\usepackage{array}
\usepackage{graphicx, subfig}
\usepackage[official]{eurosym}
\usepackage{tcolorbox}
\usepackage{tikz}
\usetikzlibrary{shadows,calc}
\graphicspath{{figures/}{plots/}}

\def\shadowshift{3pt,-3pt}
\def\shadowradius{6pt}

\colorlet{innercolor}{black!50}
\colorlet{outercolor}{gray!05}

\newcommand\drawshadow[1]{
    \begin{pgfonlayer}{shadow}
        \shade[outercolor,inner color=innercolor,outer color=outercolor] ($(#1.south west)+(\shadowshift)+(\shadowradius/2,\shadowradius/2)$) circle (\shadowradius);
        \shade[outercolor,inner color=innercolor,outer color=outercolor] ($(#1.north west)+(\shadowshift)+(\shadowradius/2,-\shadowradius/2)$) circle (\shadowradius);
        \shade[outercolor,inner color=innercolor,outer color=outercolor] ($(#1.south east)+(\shadowshift)+(-\shadowradius/2,\shadowradius/2)$) circle (\shadowradius);
        \shade[outercolor,inner color=innercolor,outer color=outercolor] ($(#1.north east)+(\shadowshift)+(-\shadowradius/2,-\shadowradius/2)$) circle (\shadowradius);
        \shade[top color=innercolor,bottom color=outercolor] ($(#1.south west)+(\shadowshift)+(\shadowradius/2,-\shadowradius/2)$) rectangle ($(#1.south east)+(\shadowshift)+(-\shadowradius/2,\shadowradius/2)$);
        \shade[left color=innercolor,right color=outercolor] ($(#1.south east)+(\shadowshift)+(-\shadowradius/2,\shadowradius/2)$) rectangle ($(#1.north east)+(\shadowshift)+(\shadowradius/2,-\shadowradius/2)$);
        \shade[bottom color=innercolor,top color=outercolor] ($(#1.north west)+(\shadowshift)+(\shadowradius/2,-\shadowradius/2)$) rectangle ($(#1.north east)+(\shadowshift)+(-\shadowradius/2,\shadowradius/2)$);
        \shade[outercolor,right color=innercolor,left color=outercolor] ($(#1.south west)+(\shadowshift)+(-\shadowradius/2,\shadowradius/2)$) rectangle ($(#1.north west)+(\shadowshift)+(\shadowradius/2,-\shadowradius/2)$);
        \filldraw ($(#1.south west)+(\shadowshift)+(\shadowradius/2,\shadowradius/2)$) rectangle ($(#1.north east)+(\shadowshift)-(\shadowradius/2,\shadowradius/2)$);
    \end{pgfonlayer}
}

\pgfdeclarelayer{shadow} 
\pgfsetlayers{shadow,main}

\newsavebox\mybox
\newlength\mylen

\newcommand\shadowimage[2][]{%
\setbox0=\hbox{\includegraphics[#1]{#2}}
\setlength\mylen{\wd0}
\ifnum\mylen<\ht0
\setlength\mylen{\ht0}
\fi
\divide \mylen by 120
\def\shadowshift{\mylen,-\mylen}
\def\shadowradius{\the\dimexpr\mylen+\mylen+\mylen\relax}
\begin{tikzpicture}
\node[anchor=south west,inner sep=0] (image) at (0,0) {\includegraphics[#1]{#2}};
\drawshadow{image}
\end{tikzpicture}}

\usepackage{listings}
\lstdefinelanguage{Keras}{
  keywords=[1]{add, compile},
  keywords=[2]{Sequential, Bidirectional, GRU, Dropout, Dense, Adam},
  keywords=[3]{input_shape, activation, loss, optimizer},
  sensitive=false,
  morestring=[b]",
}
\lstnewenvironment{kerascode}
  {\lstset{language=Keras}}
  {}

\newcommand{\compresslist}{
  \setlength{\itemsep}{1pt}
  \setlength{\parskip}{0pt}
  \setlength{\parsep}{0pt}
}

\newcolumntype{L}[1]{>{\raggedright\let\newline\\\arraybackslash\hspace{0pt}}m{#1}}
\newcolumntype{C}[1]{>{\centering\let\newline\\\arraybackslash\hspace{0pt}}m{#1}}
\newcolumntype{R}[1]{>{\raggedleft\let\newline\\\arraybackslash\hspace{0pt}}m{#1}}

\let\citet\cite

\copyrightyear{2021}
\acmYear{2021}
\setcopyright{acmlicensed}\acmConference[CHIIR '21]{Proceedings of the 2021 ACM SIGIR Conference on Human Information Interaction and Retrieval}{March 14--19, 2021}{Canberra, ACT, Australia}
\acmBooktitle{Proceedings of the 2021 ACM SIGIR Conference on Human Information Interaction and Retrieval (CHIIR '21), March 14--19, 2021, Canberra, ACT, Australia}
\acmPrice{15.00}
\acmDOI{10.1145/3406522.3446011}
\acmISBN{978-1-4503-8055-3/21/03}

\hypersetup{%
 pdftitle={\plaintitle},
 pdfauthor={\plainauthor},
 pdfauthor={\plainauthor},
 pdfkeywords={\plainkeywords},
  pdfdisplaydoctitle=true,
  bookmarksnumbered,
  pdfstartview={FitH},
  colorlinks,
  citecolor=black,
  filecolor=black,
  linkcolor=black,
  urlcolor=black,
  breaklinks,
  hypertexnames=false
}
\begin{document}
\fancyhead{}

\title{\catchytitle}
\subtitle{\plaintitle}

\author{Luis A. Leiva}
\affiliation{%
  \institution{University of Luxembourg}
  \country{Luxembourg}}
\email{name.surname@uni.lu}

\author{Ioannis Arapakis}
\affiliation{%
  \institution{Telefonica Research}
  \country{Spain}}
\email{ioannis.arapakis@telefonica.com}

\author{Costas Iordanou}
\affiliation{%
  \institution{Cyprus University of Technology}
  \country{Cyprus}}
\email{costas.jordanou@eecei.cut.ac.cy}

\begin{abstract}
This paper aims to stir debate about a disconcerting privacy issue on web browsing
that could easily emerge because of unethical practices and uncontrolled use of technology.
We demonstrate how straightforward is to capture behavioral data about the users at scale,
by unobtrusively tracking their mouse cursor movements,
and predict user's demographics information with reasonable accuracy using five lines of code.
Based on our results, we propose an adversarial method to mitigate user profiling techniques
that make use of mouse cursor tracking, such as the recurrent neural net we analyze in this paper.
We also release our data and a web browser extension that implements our adversarial method,
so that others can benefit from this work in practice.
\end{abstract}




\maketitle

\section{Introduction}

In the modern Web, privacy is becoming a rare commodity.
The recent proliferation of intrusive and privacy-invasive ads has raised serious concerns among users and industry regulatory bodies,
with initial user reaction reflected on the swift adoption of ad blocking solutions.
In fact,
most of the popular browser extensions for Mozilla Firefox
are related to ad blocking and user privacy.\footnote{\url{https://addons.mozilla.org/en-US/firefox/search/?promoted=recommended&sort=users&type=extension}}
While extensions like these have been successful in mitigating the user's exposure to web tracking,
they eventually hurt web revenue streams,
leading to the so called ``tragedy of the commons''~\cite{Cows2017},
where the common resource (user attention) will be depleted due to ad blocking.
Luckily, some effort has been put to regulate the web tracking landscape,
like self-initiatives from the ad industry that include recommendations for good practices~\cite{iab_guidelines}
and transparency tools such as AdChoices,\footnote{\url{http://youradchoices.com/}}
to help users understand why they receive specific ads.
Along the same line, privacy-preserving web browsers\footnote{See \url{https://brave.com/} and \url{https://cliqz.com/}}
allow users to have more control of their online privacy and, at the same time,
a financial incentive by the ads that they receive while surfing the web.
Moreover,
in 2018 the European Union set in place the new General Data Protection Regulation (GDPR)~\cite{gdpr} 
and the state of California in United States enforced the Consumer Privacy Act~\cite{CCPA}.
Other countries are also following the same route,
yet currently online advertising is ubiquitous.
Also, it remains the dominant monetization model on the Web,
with constantly increasing growth rates and revenues.
This has promoted advancements in user tracking and profiling technologies
that allow to serve more relevant ad content to the user, and at a higher premium, known as targeted ads or Online Behavioral Advertising~\cite{Beals_2010, Carrascosa_2015}.

Web tracking and user profiling rely on mechanisms to uniquely identify and track the user's online behavior over time,
including e.g., geolocation, visited pages, search keywords, and social network activity.
All of these in order to better understand the user intentions and interests.
However, a less known method to profile the user is by means of mouse cursor tracking.
This technology has been used successfully
to inform usability tests~\cite{Atterer:2006:KUM:1135777.1135811},
predict user engagement~\cite{Arapakis:2014:UWE:2661829.2661909}
and intent~\cite{Guo:2008:EMM:1390334.1390462, MARTINALBO2016989},
detect searcher frustration~\cite{Feild:2010:PSF:1835449.1835458},
and infer user attention to parts of a web page~\cite{Arapakis:2016:PUE:2911451.2911505},
among other tasks.
Unfortunately, because mouse cursor tracking
can be performed unobtrusively~\cite{LEIVA2015114} and at scale~\cite{Huang:2011:NCN:1978942.1979125},
it has opened the door to a brand new wave of massive tracking campaigns and companies
that hide behind laudable objectives, such as providing fine-grained, in-page analytics
(e.g., hovered and clicked items, scroll reach, speed of browsing) to the website owners.
Interestingly, by tracking the mouse cursor it is possible to profile the user demographics,
namely predicting age~\cite{Takashi2015, Yamauchi2014} and gender~\cite{Kratky:2016, Pentel2017},
a piece of valuable personal data that most users are unaware of~\cite{Carrascal13_pii}.
With this paper, we want to raise awareness about this fact
and reflect on the trade-offs between privacy and technological innovation.

We investigate and highlight privacy issues
that may emerge because of unethical practices and unregulated use of mouse cursor tracking technology.
Our contributions begin with an extensive survey of related work on privacy and security while online,
and continue with research that performed user profiling via mouse cursor data.
We then show how straightforward it is to capture behavioral data and
predict demographics information with reasonable accuracy using a few lines of code.
Based on our results, we present an adversarial method to mitigate user profiling techniques
that make use of mouse cursor tracking, such as the recurrent neural net we propose in \autoref{sssec:rnn_model}.

\section{Survey of Related Work}
\label{sec:related}

To what extent does our online activity reveal who we are?
Existing literature related to online privacy provides insights around topics such as
information leakage while surfing the web using desktop computers~\cite{WebTracking-over-years, Pujol_2015, Walls_2015, Starov_2016}
or mobile devices~\cite{Leung_2016, Mobile-Apps-NDSS2018}.
Other studies report on the entities that collect tracking data~\cite{Ads-vs-Regular-Contents, openWPM-englehardt2016census}
and how tracking data are being collected~\cite{Acar_2013, IMC2017-profit}.
Overall, this large body of work demonstrates that the digital footprints left by individuals, as they browse websites,
may help derive with alarming accuracy personally identifiable information like gender, age, location, or even political orientation.
Recent work by White et al.~\cite{White18} and Gajos et al.~\cite{Gajos20}
could detect neurodegenerative disorders from mouse cursor movements,
showing how our ``digital phenotypes'' could be used as adjunctive screening tools.
In this section, we report
current evidence on privacy risks in the online setting,
as well as predicting demographic attributes from online digital traces.

\subsection{Privacy Issues in Web Browsing}
\label{sec:privacy}

There are many ways for tracking the online activity of web users,
for example by monitoring the IP addresses
or using fingerprinting techniques~\cite{Acar_2014, Jackson:2006, MBYS11}.
However, the cookie-based approach remains to this day the dominant one, since it is fully supported by all web browsers.
With a cookie identifying the user's browser, a third party domain can track the user activity across websites
using redirection techniques or providing a free service that makes cross-domain tracking possible,
such as the Facebook `Like' button or social media sharing plugins for WordPress.
Then, by analyzing the content of the browsed websites, the tracking domain can effortlessly -- and at an unprecedented scale --
derive users' intentions and interests, alongside with other sensitive profiling information~\cite{Iordanou2019WhosTS, Ren_2016}.

There is a plethora of work that capitalize on user profiling based on how we browse.
For example, our browsing history is used
to detect targeted ads~\cite{Carrascosa_2015, MyAdChoices_Jagdish16}
or even identify which attribute or user action triggered a specific ad~\cite{xray, Wills_2012}.
Olejnik et al.~\cite{Olejnik12} noticed that, with just 4 visited websites,
it is possible to uniquely identify users in 97\% of cases.
At the same time, Web users are concerned about third-party tracking~\cite{McDonald10},
especially about location access and inferring demographics~\cite{Wills10}.
Researchers have found that people are likely to take actions to protect their privacy~\cite{Malandrino13},
including the payment of a premium fee if needed~\cite{Tsai11}.
And while some websites and tracking companies inform users about their data practices
through privacy policies and sometimes provide opt-outs~\cite{Schaub16},
these measures are insufficient.

Today, advancements in web advertising provide new opportunities to trackers and advertisers to extend their visibility.
Online ads are rendered dynamically during the load time of the browsed website,
most of the time as a result of additional (tracking) JavaScript code that is injected on the fly~\cite{Iordanou_2018}.
This opens the door to more sophisticated tracking techniques,
among which we find mouse cursor tracking to be an underestimated one.
In the following section, we highlight how this technology has been used in various scenarios
and how pervasive it has become.
In fact, today most websites include analytics scripts,
and a large number of them contains some mouse tracking script~\cite{Englehardt17}.

Users consider ad targeting useful because it highlights relevant information, but at the same time they find the underlying data collection alarming~\cite{Ur12}
and invasive~\cite{McDonald10_attitudes}.
Plane et al.~\cite{Plane17} found that users were more concerned if an ad was targeted based on demographics,
such as age, gender, or race, than based on interests.
Overall, users do not want targeted advertising
when they are made aware of the data collection methods employed by the advertisers~\cite{Turow09},
and consider targeting based on demographics to be discriminatory~\cite{Sweeney13}.
We, therefore, hypothesize that it could be possible to derive demographics information from mouse movements at scale,
and that a privacy issue may emerge if people follow unethical practices and make an uncontrolled use of the technology.

\subsection{Mouse Cursor Tracking}
\label{ssec:mouse_cursor_analysis}

What can a mouse cursor tell us more? %% VERBATIM QUESTION FROM https://dl.acm.org/doi/10.1145/634067.634234
Almost 20 years ago
Chen et al.~\cite{Chen:2001:MCT:634067.634234} raised this question
and found a relationship between gaze position and cursor position during web browsing.
Mueller and Lockerd~\cite{Mueller:2001} investigated the use of mouse tracking to visualize and (manually) infer the users' interests.
Since then, researchers have noted the utility of mouse cursor analysis
as a low-cost and scalable proxy of eye gaze~\cite{Huang:2012:USU:2207676.2208591, Navalpakkam:2013:MME:2488388.2488471}.
Several works have investigated closely the utility of mouse cursor data in web search~\cite{Arapakis:2016:PUE:2911451.2911505, 8010344, Liu:2015:DUD:2766462.2767721}
and web page usability evaluation~\cite{Arroyo:2006:CPE:1125451.1125529, Atterer:2006:KUM:1135777.1135811, Leiva:2011:RWD:2037373.2037467},
two of the most prominent use cases of this technology.
Mouse biometrics is another active research area that has recently shown how to identify an individual
by analyzing their mouse movements in controlled settings~\cite{Kratky18_biom, Lu17_biom}.

The construct of attention has nowadays become the common currency on the Web.
Objective measurements of attentional processes are increasingly sought after by the media industry,
to explain or predict user behavior.
With every click or online interaction, digital footprints are created and logged,
providing a detailed record of a person's online activity that can be used for market segmentation, targeted advertising,
but also for more privacy-invasive applications like user profiling.

Early mouse cursor tracking systems began by logging click events only (coordinates and timestamp)
and using these events to assess what information users were interested in.
However, it was soon realized that click data provide an incomplete picture of user interaction.
Click data informed researchers of a users' primary focus of attention, or their end choice.
However, a mouse click is often preceded by several interactions such as scrolling, hovers, movements, etc.
and thus can lead to a better overall understanding of the user's thought process.
This way, mouse cursor tracking systems began to incorporate such fine-grained within-page interactions to create richer user models.

In what follows, we review research efforts that have focused on mouse cursor analysis
to infer user interest, visual attention, emotions, and demographic variables like gender or age, on a desktop setting.
We thus deliberately leave out works on user profiling in mobile browsing, which fall outside the scope of this paper.

\subsubsection{Inferring User Interest}
\label{sssec:interest}

For a long time, commercial search engines have been interested in how users interact with Search Engine Result Pages (SERPs),
to anticipate better placement and allocation of ads in sponsored search or to optimize the content layout.
Early work considered simple, coarse-grained features derived from mouse cursor data
to be surrogate measurements of user interest~\cite{Claypool:2001:III:359784.359836, Shapira:2006:SUK:1141277.1141542}.
Follow-up research transitioned to more fine-grained mouse cursor features~\cite{Guo:2008:EMM:1390334.1390462, Guo:2010:RBJ:1835449.1835473}
that were shown to be more effective.
These approaches have been directed at predicting open-ended tasks
like search success~\cite{Guo:2012:PWS:2396761.2398570} or search satisfaction~\cite{Liu:2015:DUD:2766462.2767721}.
In a similar vein, Huang et al.~\cite{Huang:2012:ISM:2348283.2348313, Huang:2011:NCN:1978942.1979125}
modeled mouse cursor interactions and extended click models to compute more accurate relevance judgements of search results.
Mouse cursor position is mostly aligned to eye gaze,
especially on SERPs~\cite{Guo:2012:BDT:2187836.2187914, Speicher:2013:TPR:2505515.2505703},
and that can be used as a good proxy for predicting good and bad abandonment~\cite{Diriye:2012:LSS:2396761.2398399}.

\subsubsection{Inferring Visual Attention}
\label{sssec:attention}

Mouse cursor tracking has been also used to survey the visual focus of users in sponsored search,
thus revealing valuable -- and at the same time sensitive -- information
regarding the distribution of user attention over the various SERP components.
Despite the technical challenges that arise from this analysis,
previous work has shown the utility of mouse movement patterns
to measure within-content engagement~\cite{Arapakis:2014:UEO:3151365.3151368}
and predict reading experiences~\cite{Arapakis:2014:UWE:2661829.2661909, Hauger11}.
Lagun et al.~\cite{Lagun:2014:DCM:2556195.2556265} introduced the concept of motifs,
or frequent cursor subsequences, in the estimation of search result relevance.
Similarly, Liu et al.~\cite{Liu:2015:DUD:2766462.2767721} applied the motifs concept to SERPs
and predicted search result utility, searcher effort, and satisfaction at a search task level.
Boi et al.~\cite{Boi2016} proposed a method for predicting whether the user is looking at the content pointed by the cursor,
exploiting the mouse cursor data and a segmentation of the contents in a web page.
Lastly, Arapakis et al.~\cite{Arapakis:2016:PUE:2911451.2911505, Arapakis20_ppa}
investigated user engagement with direct displays on SERPs
and provided further evidence that supports the utility of mouse cursor data
for measuring user attention at a display-level granularity.

\subsubsection{Inferring Emotional State}
\label{sssec:emotion}

Although the connection between mouse cursor movements and the underlying psychological states has been a topic of research
since the early 90s~\cite{Accot1997, Card:1987},
some studies have investigated the utility of mouse cursor data for predicting the user's emotional state.
For example, Zimmermann et al.~\cite{Zimmermann2003} investigated the effect of induced affective states
on the motor-behavior of online shoppers and found that the total duration of mouse cursor movements
and the number of velocity changes were associated to the experienced arousal.
Kaklauskas et al.~\cite{Kaklauskas2009}
created a system that extracts physiological and motor-control parameters
from mouse cursor interactions and then triangulated those with psychological data taken from self-reports,
to analyse correlations with users' emotional state and labour productivity.
In a similar line, Azcarraga et al.~\cite{Azcarraga:2012} combined electroencephalography signals and mouse cursor interactions
to predict self-reported emotions like frustration, interest, confidence and excitement.
Yamauchi et al.~\cite{Yamauchi:2013} studied the relationship between mouse cursor trajectories and generalized anxiety in human subjects.
Lastly, Kapoor et al.~\cite{Kapoor:2007} predicted whether a user experiences frustration, using an array of affective-aware sensors.

\subsubsection{Inferring Demographics}
\label{sssec:demographic_attributes}

Yamauchi et al.~\cite{Yamauchi2014} examined the extent to which mouse cursor movements can help identify the gender
and the experienced feelings of users who were watching short film clips.
Although this work provides early evidence on the utility of mouse cursor data for advanced online user profiling,
it suffers from certain limitations that we address in this work.
First, the experimental setting has limited generalizability,
since the adopted perception task is not very well connected
to typical activities that users perform online, such as web search.
Second, the data used in their predictive modeling task include multiple samples per participant
randomly assigned to the training and test data partitions,
hence there may be information leakage that artificially inflated model performance.
In our analysis, we limit the training samples to exactly one mouse cursor trajectory per participant
and test our models on unseen individuals.

Kratky et al.~\cite{Kratky:2016} recorded mouse cursor movements in an e-commerce website
and engineered a set of meta-features to predict the user gender and age group.
Their classifier was trained on several days of data per participant.
Although the training and test collections had disjoint sets of participants,
it was stated that the reported results were overly optimistic
since researchers could not verify their ground-truth data~\cite{Kratky:2016}.
In contrast, as discussed later,
our dataset was collected from high-quality crowdworkers
so we are confident that the ground-truth information is correct.

In a similar vein, Pentel et al.~\cite{Pentel2017} used data from six different external sources,
including e.g., keystroke data and feedback questionnaires,
and handcrafted features proposed in earlier works~\cite{Claypool:2001:III:359784.359836, Diriye:2012:LSS:2396761.2398399, Shapira:2006:SUK:1141277.1141542}
to train predictive models that could identify the users' age and gender.
However, because their approach relies mainly on ad-hoc data,
it is less scalable and more difficult to implement than the approach we propose in this paper,
which takes as input \emph{raw} mouse cursor data.
Moreover, Pentel et al. reported optimistic performance scores,
which may be due to information leakage across data partitions,
and omit important classification metrics such as precision, recall, and AUC.
To account for their modeling approach, as well as that proposed by Kratky et al.~\cite{Kratky:2016},
we implement the same classifier and test it in our setting (see~\autoref{ssec:results}).

\subsection{Summary}

Websites can infer fine-grained information about the users by tracking their mouse cursor activity.
Tracking where exactly on the page a user's mouse cursor hovers or clicks provides a surrogate signal for gaze fixation,
and therefore reveals the focus of attention, which can be used to learn the users' \emph{latent} interests.
However, the research literature on mouse cursor tracking
has pointed out far more advanced and creative use cases for this technology.
The above studies demonstrate that certain cognitive and motor control mechanisms are embodied
and reflected, to some extent, in our mouse cursor movements and online interactions.
In other words, mouse cursor movements can disclose sensitive information
that may be employed for advanced user profiling,
such as the identification of demographics, personality traits, and browsing intent.
For brevity's sake, in the remainder of this paper we will focus on predicting demographics (age and gender) from mouse cursor movements,
but we argue that other, potentially sensitive information may remain vulnerable
and could be exposed if appropriate mechanisms are put into place.

\section{Study}
\label{sec:user_study}
We ran an online user study a few years ago 
that reproduced the setting of a \emph{sponsored search} task~\cite{Leiva20_att}.
In order to make this paper self-contained, we will describe here the data collection procedure
in enough detail to allow reproducibility of our work,
nevertheless the reader may consult our reference paper~\cite{Leiva20_att} for more details.

Sponsored search provides the necessary revenue streams
to commercial web search engines\footnote{\url{https://searchengineland.com/google-search-ad-revenues-271188}}
and it is critical to the success of many websites~\cite{Jansen2008SponsoredSA}.
Commercial web search engines resort to various tracking techniques to monitor their users' search activity,
including mouse cursor tracking~\cite{Diaz13, Huang:2012:USU:2207676.2208591, Huang:2011:NCN:1978942.1979125},
and use that information to offer item recommendations~\cite{Speicher:2013:TPR:2505515.2505703},
targeted advertising~\cite{Bashir_2016},
or simply sell it to third parties~\cite{Morey15}.

With this work, we critique the use of mouse cursor tracking technology,
highlighting possible implications for the future of the online advertising industry. %sponsored search process.
More specifically, our user study allowed us to capture in a non-invasive manner
the mouse cursor interactions of users who performed simple web search tasks.
The collected mouse cursor data was then used to benchmark state-of-the-art
machine learning models' capacity to infer users' demographic attributes.

\subsection{Design}
\label{ssec:design}

\begin{figure}[!tpb]
    \centering
    \shadowimage[clip, width=0.75\linewidth]{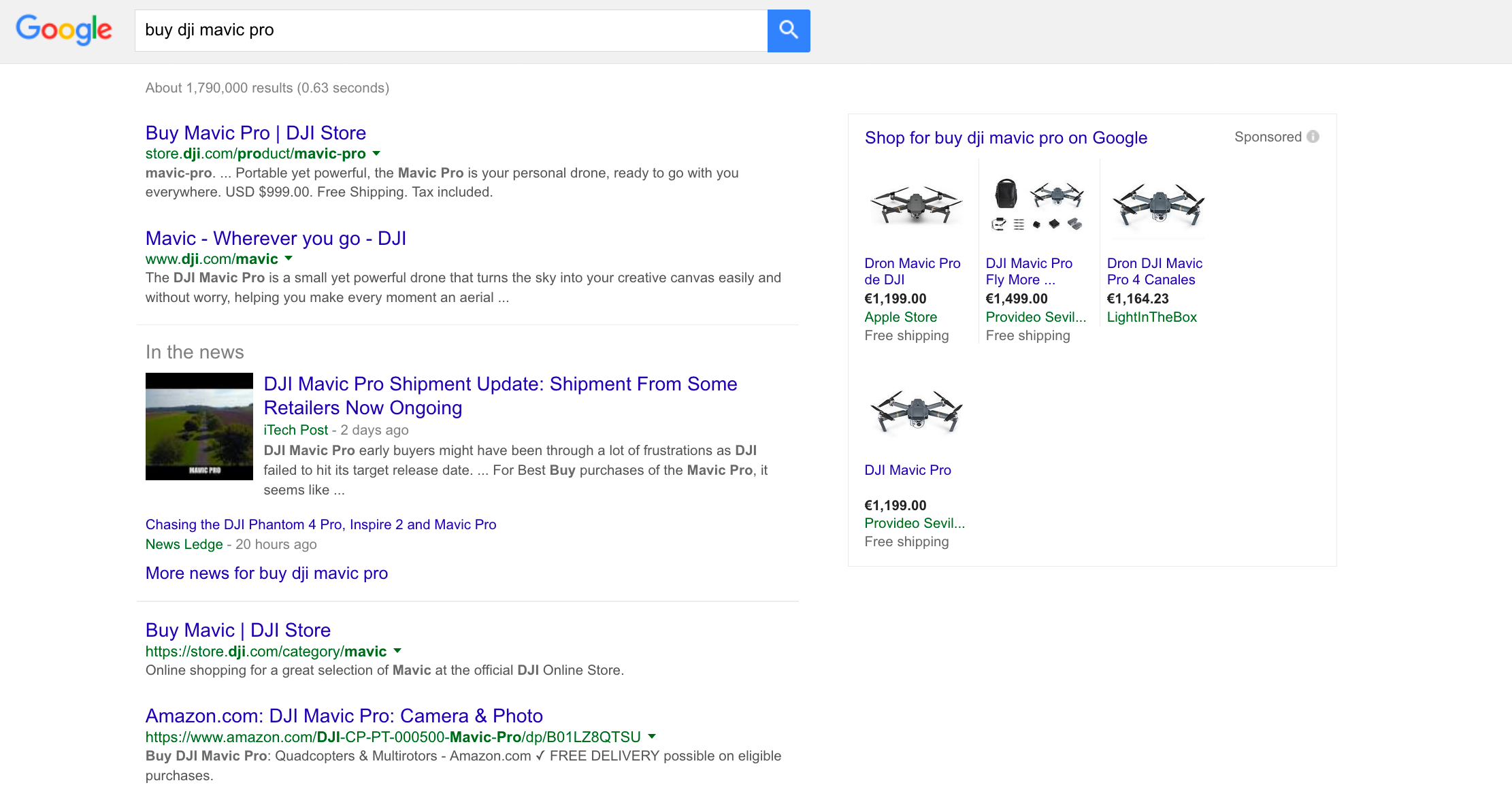}
  \caption{
    Example of a Google SERP used in the online study.
  }
  \label{fig:serp}
\end{figure}

Our experiment,  which was approved by a team of legal experts, 
consisted of a brief transactional search task~\cite{Broder2002} that was completed once per participant.
Participants were presented with a simulated information need
that explained that they were interested in purchasing a present for them or a friend,
and were asked to use Google Search to find something appealing.
Each participant was provided with a predefined search query and the corresponding SERP (see~\autoref{fig:serp})
and were asked to click on any element of the SERP that answered it best.
This way, we ensured that participants interacted with the same pool of web search queries
and avoided any unaccounted systematic bias due to query quality variation.
Overall, the search task consisted of three parts:
(1)~pre-task guidelines, (2)~the web search task and (3)~a post-task questionnaire.

The search queries (\autoref{sssec:search_query_sample}),
which were all picked from a pool of popular queries in Google Search,
were randomly distributed among our participants.
The corresponding SERPs appeared all in English and were scraped for later instrumentation,
simulating thus a website owner who wishes to track their users' every move.

Participants accessed the instrumented SERPs through a dedicated server that did not alter the look and feel of the original SERPs.
This allowed us to capture fine-grained user interactions
while ensuring that the content of the SERPs remained consistent with the original version.
Each participant was allowed to perform the search task \emph{only once}
to avoid introducing possible carry over effects
and, thus, altering their browsing behavior in subsequent search tasks.

\subsection{Apparatus}
\label{ssec:apparatus}

\subsubsection{Search Query Sample}
\label{sssec:search_query_sample}

Starting from Google Trends,\footnote{\url{https://trends.google.com/trends/}}
we selected a subset of the Top Categories and Shopping Categories that were suitable representatives of transactional tasks~\cite{Broder2002} i.e. categories that broadly express the intent of performing some web-mediated activity or transaction, like shopping or finding a service. Then, we extracted the top search queries issued in the US during the last 12 months and further narrowed down our search query collection to the 150 most popular search queries.
Using this final selection of search queries, we produced the static version of the corresponding Google SERPs
and injected JavaScript code (see next section) that allowed us to capture all client-side user interactions.

\subsubsection{Mouse Cursor Tracking}
\label{sssec:mouse_cursor_tracking}

As previously stated, all SERPs were downloaded and instrumented with custom JavaScript code.
This way, we could automatically insert mouse tracking code and log cursor movements, hovers, and associated metadata.
For this, we used \textsc{EvTrack},\footnote{\url{https://github.com/luileito/evtrack}}
an open source JavaScript event tracking library
derived from the smt2$\epsilon$ mouse tracking system~\cite{Leiva:2013:WBB:2540635.2529996}.

We captured \texttt{\small{mousemove}} events via event polling,
every 150\,ms and all the other browser events (e.g., \texttt{\small{load}}, \texttt{\small{click}}, \texttt{\small{scroll}}) via event listeners. Whenever an event was recorded, we logged the following information:
mouse cursor position ($x$ and $y$ coordinates), timestamp, event name,
XPath of the DOM element that relates to the event, and the DOM element attributes (if any).
\textsc{EvTrack} has no dependencies and works in every major browser so, upon download,
it is ready to use; i.e. no tooling or build pipeline is needed.
This ease of use reveals how straightforward is to add mouse tracking capabilities to websites.

\subsubsection{Questionnaire}
\label{sssec:questionnaire}

In addition to the mouse cursor data, we gathered ground-truth information about the users
through an online questionnaire that was administered at post-task.
The questions included in the questionnaire were forced-choice type
and allowed multi-point response options.
The questionnaire comprised the following questions:
\begin{enumerate}[leftmargin=*]
\compresslist
\item \emph{What is your gender?} [Male, Female, Prefer not to say]
\item \emph{What is your age group?} [18--23, 24--29, ..., 60--65, +66, Prefer not to say]
\item \emph{What is your native language?} [Pull-down list, Prefer not to say]
\end{enumerate}

\subsection{Participants}
\label{ssec:participants}

We recruited participants from the \textsc{Figure Eight} crowdsourcing platform.\footnote{\url{https://www.figure-eight.com}}
They were of mixed nationality and had diverse educational backgrounds. All participants were proficient in English and were experienced (Level 3) contributors, i.e. they had a track record of successfully completed tasks and of a different variety, thus being considered very reliable contributors.

\subsection{Procedure}
\label{ssec:procedure}

Participants were instructed to read carefully the terms and conditions of the study which,
among other things, informed them that they should perform the task from a desktop or laptop computer
using a computer mouse (and refrain from using a touchpad, tablet, or mobile device)
and that their browsing activity would be logged.
Moreover, participants consented to share their browsing data and their questionnaire responses for later analysis.

Participants were asked to act naturally and choose anything that would best answer a given search query,
since all ``clickable'' elements (e.g., result links, images, etc.) on the SERP were considered valid answers.
The instructions were followed by a brief search task description like ``\emph{You want to buy $<$noun$>$ (for you or someone else as a gift) and you have submitted the search query $<$noun$>$ to Google Search. Please browse the search results page and click on the element that you would normally select under this scenario.}''

The SERPs were randomly assigned to the participants
and each participant could take the study only once (see~`\nameref{ssec:design}' section).
Participants were allowed as much time as they needed to examine the SERP and proceed with the search task,
which concluded upon selecting any of the ``clickable'' elements on the SERP.
At the end of the task, participants were asked to complete the post-task questionnaire.
The payment for participation was \$0.20 and the study took 0.83 minutes on average to complete (Mdn=0.37, SD=2.3)
which roughly amounts to a 14\$/h wage.
Participants could also opt-out at any moment, in which case they were not compensated.

\subsection{Dataset}
\label{ssec:dataset}

After excluding the users who did not provide their demographic information
(see `\nameref{sssec:questionnaire}' section)
and had few mouse movement data
(less than ten mouse coordinates, which corresponds roughly to two seconds of user interaction data),
we concluded on a set of $1,467$ search sessions.
The average mouse cursor trajectory length was 25.2 coordinates (SD=18.7, min=11, max=221).
Next, our dataset was divided into a 90:10 training-test split;
i.e., 90\% of the data is used for model training
and the remaining 10\% of the data is used for testing.
Our raw dataset is publicly available.\footnote{\url{https://gitlab.com/iarapakis/the-attentive-cursor-dataset}}

\subsection{Machine Learning Models}
\label{ssec:models}

The focus of these experiments is demonstrating how feasible it is to implement a user profiling mechanism
by relying on current machine learning techniques and easily acquired mouse cursor data.
Therefore, for the sake of simplicity, we assume gender and age classification to be a two-class problem,
i.e. a user is classified as `male' or `female' and as `young' or `adult'.
We note that age could be framed as a regression problem,
however marketing companies care more about market segmentation (i.e. fitting customers into target groups)
rather than predicting a particular age~\cite{Oh02, Schewe00}.

\subsubsection{Baseline Models}
\label{sssec:baseline}

We replicate the random forest (RF) classifier proposed in recent work~\cite{Pentel2017, Yamauchi2014},
which is an effective ensemble method that allows for a reliable performance assessment. Furthermore, we engineer a series of features (e.g., speed, acceleration, angle, traversed distance, hovers, clicks)
and aggregate functions (e.g., min, max, mean and standard deviation) derived from the mouse cursor data, as reported in previous work~\cite{Yamauchi2014, Pentel2017} (170 features).
Then, we exclude the highly correlated ($r \ge .80$, $p < .05$) and linearly dependent features from our feature set
and normalize the values for all features in the $[0,1]$ range,
so that feature values that fall in greater numeric ranges would not dominate those in smaller numeric ranges.
In total, the RF model uses 52 mouse cursor features for classification.
As a last step, we determine via grid search on a held-out set (comprising 10\% of the training data)
the optimal hyper-parameter values (number of trees, number of features, $\epsilon$-threshold, minimum size of terminal nodes, maximum number of terminal nodes) and evaluate the performance of the RF model against the test set.

We also implement a ZeroR classifier, also known as 0-R (zero rule),
which simply predicts the majority class.
The ZeroR will always output the same target value and does not use any input features, hence its name.
Despite its simplicity and lack of discriminative power,
this classifier is very useful for determining the baseline performance,
as a benchmark for other classification methods like the ones we used in these experiments.

\begin{figure*}[!ht]
  \centering
  \def\w{0.23\textwidth}
  \includegraphics[width=\w]{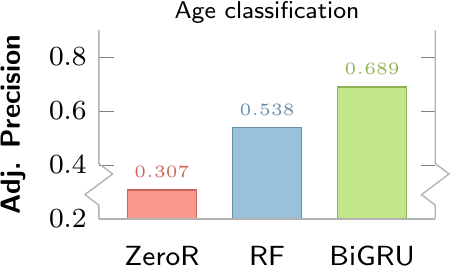}
  \hfill
  \includegraphics[width=\w]{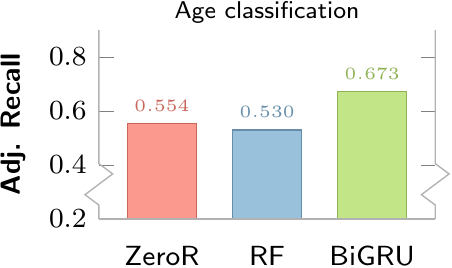}
  \hfill
  \includegraphics[width=\w]{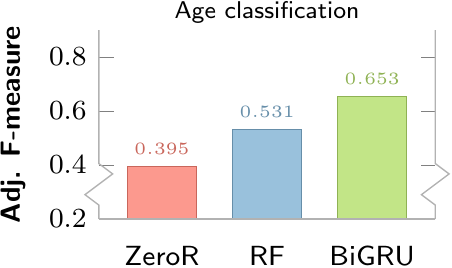}
  \hfill
  \includegraphics[width=\w]{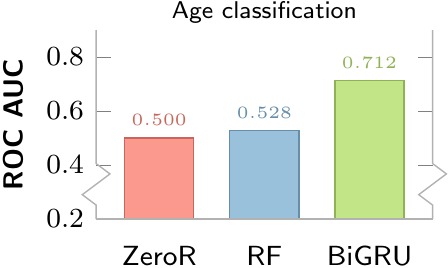}
  \caption{Age classification results.
    All metrics are weighted by class distribution.}
  \label{fig:results-age}
\end{figure*}

\begin{figure*}[!ht]
  \centering
  \def\w{0.23\textwidth}
  \includegraphics[width=\w]{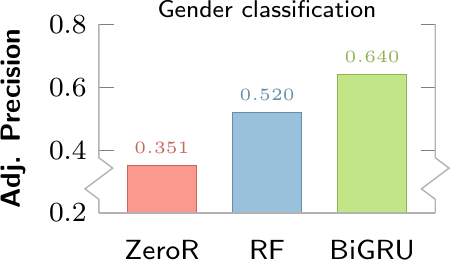}
  \hfill
  \includegraphics[width=\w]{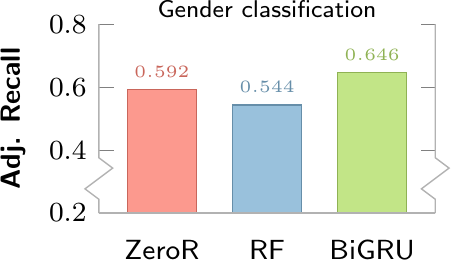}
  \hfill
  \includegraphics[width=\w]{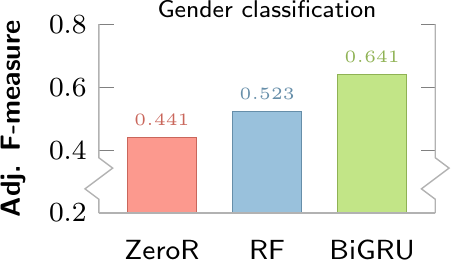}
  \hfill
  \includegraphics[width=\w]{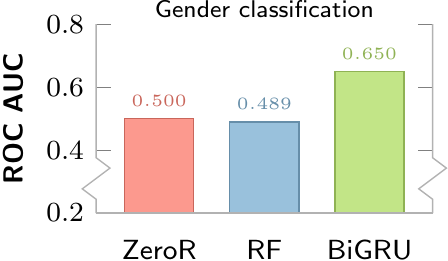}
  \caption{Gender classification results.
    All metrics are weighted by class distribution.}
  \label{fig:results-gender}
\end{figure*}

\subsubsection{Recurrent Neural Network Models}
\label{sssec:rnn_model}

Creating a competent feature-based classifier like the RF previously described,
as noted, demands significant effort and time because of the hyperparameter fine-tuning and, above all, the feature engineering process.
Indeed, feature engineering requires domain expertise to derive features with sufficient discriminative power.
With neural networks, however, feature engineering is automatically performed by the network itself.
Together with the availability of state-of-the-art deep learning libraries such as Tensorflow, Keras, PyTorch, or MXNet, it has become increasingly easy to implement a competent classifier with few lines of code
(see our implementation in~\autoref{fig:code})
and, hence, the purpose of this paper.

Since mouse movements are of sequential nature,
we test a particular type of recurrent neural networks (RNNs)
that is effective at modeling time series,
where each data point in the sequence can be assumed to be dependent on the previous one.
Concretely, the model uses Gated Recurrent Unit (GRU) memory,
which is a simplification of the popular long short-term memory.
We use the bidirectional variant (BiGRU) since a major issue with all RNNs
is that they can only learn representations from \emph{previous} time steps.
However, sometimes we have to learn representations from \emph{future} time steps
to better understand the context and thus eliminate potential ambiguities.

Our BiGRU takes as input a \emph{raw} sequence of mouse cursor positions and time offsets,
which can be seen as a multivariate time series of three-dimensional data points.
Because each mouse sequence has a different length,
all sequences are padded to a fixed length of 100 timesteps,
which corresponds roughly to the mean sequence length observed in our dataset plus three standard deviations.
The input layer of our RNN model has 100 neurons (one neuron per timestep).
The hidden layer is the forward-backward recurrent block (BiGRU) with 64 output units,
using hyperbolic tangent activation and sigmoid activation in the recurrent step.
We add a dropout layer with drop rate $q=0.25$ for regularization,
followed by a fully-connected (FC) layer of 1 output unit using sigmoid activation.
The model outputs a probability prediction $p$ of the user's gender or age,
where $p>.5$ indicates that the user belongs to the majority class
(in our data, `male' and `young' are the majority classes).

We use the popular Adam optimizer (stochastic gradient descent with momentum)
with learning rate $\eta=0.0005$ and decay rates $\beta_1=0.9$ and $\beta_2=0.999$.
This model, including all the settings described above,
is implemented in five lines of Python code; see \autoref{fig:code}.
We train this model with a batch size of 32 sequences and up to 400 epochs,
with early stopping of 40 epochs to prevent overfitting.

\section{User Profiling Experiments}
\label{ssec:results}

We report the weighted Precision, Recall, and F-measure (F1 score),
according to the target class distributions in each case.
In addition, we provide the Area Under the ROC curve (AUC), to highlight the discriminative power of each classifier.
Finally, we remind the reader that the focus of this paper is not on attaining state-of-the-art performance but rather on demonstrating that it is feasible to implement a fairly competent user profiling mechanism
by relying on current machine learning techniques and easily acquired mouse cursor data.

\subsubsection{Age Classification}
\label{sssec:age}

Prior work has linked age with motor control and pointing performance in tasks
that involve the use of a computer mouse~\cite{Bohan1998, HSU1999461, Jastrzembski2003InputDF, LINDBERG2006170, Smith1999, Walker1997}.
Overall, ageing is marked by a decline in motor control abilities,
therefore it is expected to affect the users' pointing performance and, by extension, how they move the computer mouse.
For example, Smith et al.~\cite{Smith1999} observed that older people incurred in longer mouse movement times,
more sub-movements, and more pointing errors than the young.
These findings underline potential age effects on the way a mouse device is used in an online search task.

\autoref{fig:results-age} shows the performance results for the classification task that targets user age.
Here, we divide our users into two age groups (``18--35'' and ``36--66''),
in line with previous work~\cite{Kratky:2016, Pentel2017} that applied a comparable binary split on their user sample.
While the RF model achieved an F-measure of 0.531 and an AUC of 0.528,
the BiGRU outperformed its peers with an F-measure of 0.653 and an AUC of 0.712.
Furthermore, we ran pairwise comparisons of proportions
(Bonferroni-Holm corrected, to guard against over-testing the data)
and observed statistically significant differences for all metrics
when comparing the BiGRU against the other classifiers ($p<.01$).

We also note here that the performance of the RF model is much smaller than
what researchers have reported in previous work~\cite{Kratky:2016, Pentel2017},
whereas the simple implementation of our BiGRU model,
only with raw mouse movements as input (spatial coordinates and time offsets)
and five lines of code (\autoref{fig:code}),
validates the need to raise further awareness about the potential threats of mouse cursor tracking to online privacy.

\subsubsection{Gender Classification}
\label{sssec:gender}

Prior research noted sensory-motor differences due to gender~\cite{Chen2008TheEO, Landauer1981, Takashi2015},
such as significant variation in the cursor movement distance, pointing time, and cursor patterns.
The cause of these variations has been attributed to gender-based differences in how users move a mouse cursor
or to different cognitive mechanisms (perceptual and spatial processes) involved in motor control.

\autoref{fig:results-gender} shows the performance results for the classification task that targets user gender.
Again, the BiGRU model outperformed its peers.
More specifically, the RF model achieved an F-measure of 0.523 and an AUC of 0.489,
while the BiGRU achieved an F-measure of 0.641 and an AUC of 0.650.
The pairwise comparisons of proportions (Bonferroni-Holm corrected) revealed
statistically significant differences for all metrics except Recall
when comparing the BiGRU against the other classifiers ($p<.01$). Although these results might not be as impressive as those pertaining age classification, they clearly deviate from random classification and definitely call for attention to the potential implications for e-privacy.

In addition, we observe that,
unlike the optimistic results reported by others~\cite{Pentel2017, Yamauchi2014},
the same RF model performed worse on our data,
possibly due to the more challenging nature of the task.
More importantly, we have shown that a shallow BiGRU model can outperform
a predictive model that relies on a barrage of elaborate features,
by using exclusively as input unprocessed mouse cursor movements, which are easy to acquire unobtrusively and at scale.
Hence, sensitive information may be exploited by anyone who has access to a simple profiling technology, such as the one demonstrated in this paper.

\section{Profiling Prevention Experiments}
\begin{figure}[!ht]
    \centering
    \includegraphics[width=0.9\linewidth]{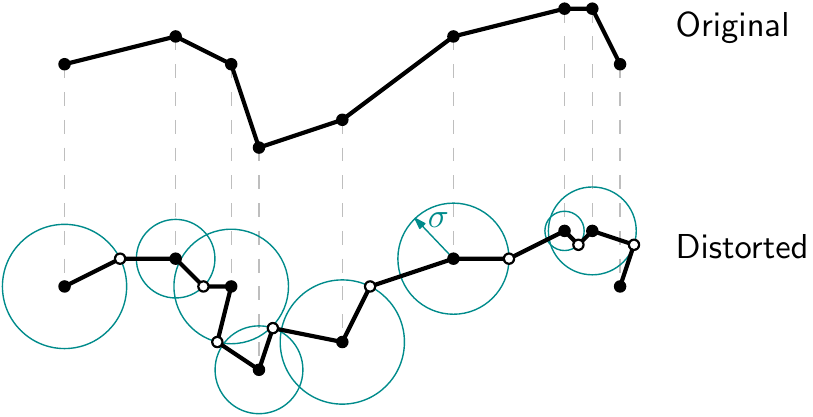}
    \caption{Adversarial noise example.
        We add an intermediate coordinate programmatically between two consecutive coordinates
        that is $\sigma$\,px away from the current position.}
    \label{fig:noise}
\end{figure}

Now that we know that it is possible to easily infer user demographics from mouse cursor movements,
we propose an adversarial method to modify the user's movements
in such a way that the resulting trajectory cannot disclose age and gender information.
The method is illustrated in \autoref{fig:noise}:
Whenever a \texttt{\small{mousemove}} event $e_t$ happens at time $t$,
we insert another \texttt{\small{mousemove}} event programmatically
$e'_{t} \sim \mathcal{N}(0, \sigma)$
which is within a $\sigma$ radius away from the original coordinate.
This additive Gaussian noise is also applied to the time offsets,
to ensure that the distorted trajectory has no duplicated times.

The amount of adversarial noise applied to each programmatic event ranges randomly from 0 to $\sigma$.
We ensure that distorted points (both coordinates and time)
are always positive values, in line with regular mouse movement data.
In this experiment, we study the impact of $\sigma$ in classification performance.
Theoretically, a random classifier should achieve an AUC score of $0.5$
for a two-class classification problem.
Therefore, we expect to see a degradation in classification performance
with regard to the previous experiments.
Given that the BiGRU outperformed the RF model
and relies only on raw mouse movements,
we only challenge the BiGRU model in these experiments.

\begin{figure}[!ht]
    \centering
    \def\w{\linewidth}
    \includegraphics[width=\w]{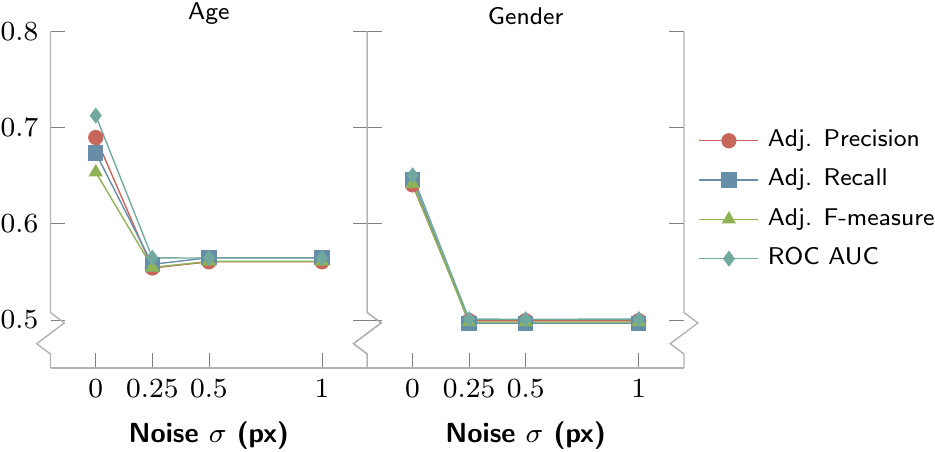}
    \caption{Degradation of classification results (weighted by class distribution)
        after introducing adversarial noise.}
    \label{tbl:results_with_noise}
\end{figure}

As observed in \autoref{tbl:results_with_noise}, using a radius of $\sigma=0.25$\,px
is enough to degrade the performance of the BiGRU model,
which begins to behave as a random classifier (AUC $\approx 0.5$)
of both age and gender.
The differences between the original mouse data and the degraded versions
are statistically significant at the $p < .001$ level.
Hence, this experiment justifies further our proposal to prevent user profiling techniques
that exploit mouse cursor movements data.
We argue that by using this adversarial noise,
a mouse movement trajectory would become ``illegible'' to any machine learning model
trying to classify the user's age or gender.
This experiment is inline with previous research that reported
very small perturbations to cause a significant performance degradation~\cite{Prakash18_px}.
Indeed, by scrambling both spatial and temporal information from a mouse cursor trajectory
we are effectively signaling a seemingly arbitrary and jittery movement.

To validate further the validity of this adversarial noise,
we re-trained our BiGRU model with distorted mouse data,
using the same configuration from our previous experiments (\autoref{sssec:rnn_model}).
Now each mouse sequence is distorted according to a random uniform distribution $\sigma \sim \mathcal{U}(0,1)$,
which means that some coordinates are preserved ($\sigma=0$)
whereas others are more distorted ($\sigma=1$).
The results are shown in \autoref{tbl:adversarial-retrain}.
As can be observed, the model achieves worse performance
than the model trained on the original, non-distorted data,
sometimes by a large margin.
This was especially so for the age classifier.
We conclude that the proposed adversarial noise technique is a robust
countermeasure against the neural net used as profiling mechanism via mouse cursor tracking.

\begin{table*}[!ht]
  \begin{tabular}{l *8r}
  \toprule
  \textbf{Demographics} & \multicolumn{2}{c}{\textbf{Adj. Precision}}
                        & \multicolumn{2}{c}{\textbf{Adj. Recall}}
                        & \multicolumn{2}{c}{\textbf{Adj. F-measure}}
                        & \multicolumn{2}{c}{\textbf{ROC AUC}} \\
  \midrule
  Age    & 0.6301 & \textcolor{gray}{$\downarrow$8.55\%}
         & 0.5986 & \textcolor{gray}{$\downarrow$11.05\%}
         & 0.5328 & \textcolor{gray}{$\downarrow$18.4\%}
         & 0.6074 & \textcolor{gray}{$\downarrow$14.7\%} \\ 

  Gender & 0.6429 & \textcolor{gray}{$\uparrow$0.45\%}
         & 0.6463 & \textcolor{gray}{$\downarrow$0.04\%}
         & 0.6133 & \textcolor{gray}{$\downarrow$4.32\%}
         & 0.6308 & \textcolor{gray}{$\downarrow$2.95\%} \\ 
  \bottomrule
  \end{tabular}
  \caption{Test performance of our BiGRU when trained using adversarial noise \bm{$\sigma \sim \mathcal{U}(0,1)$}.
    We show the performance degradation in parentheses,
    as a the percentage variation w.r.t the same classifier trained on non-distorted data.
  }
  \label{tbl:adversarial-retrain}
\end{table*}

We have implemented this adversarial method in a Chrome extension (see `\nameref{apx:resources}' section)
that allows to configure the adversarial noise level (random uniform by default)
and the number of \texttt{\small{mousemove}} events (one coordinate by default) to be added programmatically.
The idea of providing these ``sane defaults'' is to avoid making the mouse movements too distorted
so as they can go unnoticed by a machine learner trying to distinguish between human and fake movements.

% ========================================================
\section{Discussion}
% ========================================================

The rapid growth of online advertising has spurred the demand for effective, but also at times privacy-invasive
user profiling technologies that allow to deliver more relevant ad content to the user.
Of course, user profiling is not a bad thing per se, 
since it allows to deliver more relevant ad content to the user. 
However, as targeted ads are believed to produce increased revenues, 
various intermediary companies (such as ad platforms and ad exchanges) are tracking users at scale, 
and often in an unregulated manner, which has resulted in a privacy nightmare~\cite{MyAdChoices_Jagdish16}.
Advertising is currently the main business model of ``free'' content and services on the Web,
though if something is free it usually means that the user is the product.
This dystopian reality evokes a disconcerting dichotomy between, on the one hand,
having to accept a digital life with no privacy and, on the other hand,
retaining our privacy by being off the digital life.

Privacy is essential for the citizens of both the physical and the digital world.
But also privacy is constantly being juxtaposed with competing goods and interests,
balanced against disparate needs and demands~\cite{Pozen16}.
More importantly, the loss of privacy translates into a loss of freedom.
In other words, freedom of expression is threatened by the surveillance of our digital traces;
thought patterns and intentions can be extrapolated from website visits (rightly or wrongly),
and the knowledge that we are being surveyed can make one less likely to research a particular
topic.\footnote{\url{https://robindoherty.com/2016/01/06/nothing-to-hide.html}}
And even efforts to regulate the web tracking landscape, such as the DNT Header,
require good faith cooperation from the parties at the other end of the web connection, which is not always guaranteed.
In fact, most tracking domains and ad platforms are unlikely to ignore user tracking,
because they make their business out of these data.
Hence, the Web privacy problem is a fallout of rapid and uncontrolled growth in technology,
mainly driven by a lack of transparency, control, and difficulty to understand e-privacy implications.

\subsection{Implications and Outlook}

Mouse cursor tracking is very difficult to avoid while browsing the Web today.
Our mouse movements can be tracked silently at scale, in Incognito mode,
and even without JavaScript enabled~\cite{Huang10_css}. Being a low-cost and reasonable proxy of visual attention,
mouse cursor tracking cannot be discounted from being a modern ``Trojan horse''.
Our analysis corroborates this account and demonstrates that,
through a very simple machine learning implementation,
we can infer people's gender or age inadvertently. We do not argue, however, that mouse tracking should be removed from any website, as it may be a valuable data source for various application scenarios.
Rather, we find it disconcerting that currently there is no way for end-users to opt-out easily.

The work presented here serves a dual purpose.
First, it aims at raising awareness on the emerging privacy threats in the online world
and exposes some of the unaccounted ---yet sizeable risks--- of tracking technologies.
Even when dealing with a seemingly harmless browser activity logging practice,
such as the collection of mouse cursor coordinates,
a third party can mine this data to uncover personally identifiable information about the users.
Second, with this work we intend to give control back to the users over their (mouse) data.
We are certainly not the first ones in aiming at this goal, though,
to the best of our knowledge, our proposed adversarial noise technique
is the first countermeasure against user profiling based on mouse cursor tracking.
We note, however, that our method will not prevent ``any and all'' profiling techniques within mouse tracking,
as the field of adversarial machine learning progresses rapidly
and new counter-countermeasures are likely to appear in the future.

Researchers have proposed restricting access to only the features
which are necessary for delivering a desired functionality~\cite{Snyder17},
enforcing thus a principle of least privilege~\cite{Denning76}.
Google Chrome have recently announced a new technical proposal named ``privacy budget''
that could restore the balance between user privacy and ad targeting on the Web.\footnote{%
\url{https://blog.chromium.org/2019/08/potential-uses-for-privacy-sandbox.html}
}
With a privacy budget, websites could call APIs until those calls have revealed enough information,
to narrow a user down to a group sufficiently large enough to maintain anonymity.
After that, any further attempts to call such APIs would cause the browser to intervene and block further calls.
Under this scenario, we could imagine a user-configurable privacy budget for mouse tracking data,
though a set of sane defaults should be provided by browser vendors.
For example, if a JavaScript function is listening to the \texttt{onmousemove} event more than $N$ seconds in a row,
the browser would block further calls to the listener function. We would also recommend browser vendors to list what sensitive APIs a website is using, just like they do currently to inform about SSL certificates, for example,
or even ask for explicit consent to the user when the website requires access to such sensitive APIs,
similar to app permission requests in mobile devices.

Finally, we hope that this work will motivate further research
to counterbalance initiatives on developing privacy-invasive user profiling technologies
by delivering techniques that can preserve user anonymity and protect personally identifiable information.
This work also highlights the need for more transparency and privacy-aware tools.
We believe that users should be tracked, if at all, by category instead of individually.
While some advertisers do care about organizing users into general groups,
others aim at creating detailed individual profiles,
which should not happen without explicit user's consent.

\subsection{Limitations and Future Work}

We have analyzed movements generated by a computer mouse and so the proposed method
is not expected to work ``as is'' on touch-capable devices, such as tablets or smartphones.
However, user engagement is still higher on desktop than on mobile~\cite{Arapakis20_ppa},
which means more profitability for advertisers. Nonetheless, this presents an opportunity to extend our work and account for touch-based interactions such as, for example, tracking zoom/pinch gestures and scroll activity
instead of the mouse cursor position~\cite{Guo:2013:MTI:2484028.2484100}.

We have shown that it is possible to detect user demographics with reasonable accuracy.
More importantly, we have shown that is possible to do so unobtrusively and at scale,
by relying only on sequences of raw mouse cursor data.
However, since the focus of our work in not on attaining state-of-the-art performance,
there is still room to benchmark further the capabilities of such user profiling technologies
and uncover additional vulnerabilities in the data.
For example, stacking more recurrent layers (deeper model),
increasing the number of hidden neurons (wider model),
or using data augmentation techniques. Even non-sequential models are also possible to analyze mouse cursor data~\cite{Arapakis20_mtdl}.

Finally, we acknowledge a limitation of our adversarial noise technique.
The W3C consortium introduced the concept of ``trusted events'',\footnote{\url{https://www.w3.org/TR/uievents/\#trusted-events}} to help developers differentiate between events triggered by a genuine user interaction
and those triggered programmatically, e.g., by a 3rd party script.
Our Chrome extension adds mouse cursor distortions programmatically via JavaScript,
therefore those events are considered untrusted,
although currently none of the major mouse tracking companies
filter out untrusted DOM events.\footnote{See e.g., \url{https://easylist.to/easylist/easyprivacy.txt}} It is a matter of time, however, for companies to catch up and update their tracking technology. Therefore, in future work we will release a program that runs at the Operating System level\footnote{\url{https://github.com/jordansissel/xdotool}} and thus can trigger mouse events that are seen as trusted by the web browser.

\section{Conclusion}

It is possible to infer user demographics unobtrusively and at scale
with reasonable accuracy, using an off-the-shelf recurrent neural network
that takes as input raw mouse movements.
Previous attempts have relied on expert knowledge in machine learning techniques and feature engineering methods.
Therefore, we noticed an unprecedented low entry barrier for webmasters
interested in profiling the user on their websites with no effort,
highlighting thus a disconcerting privacy issue. We have proposed an adversarial noise method to mitigate such user profiling
techniques that make use of mouse cursor tracking to predict demographic variables such as gender and age,
so that users interested in preserving their privacy can do so with no effort too. It is our hope that this paper will raise awareness among the research community
about how easy the task of profiling users on the Web has become,
including mouse cursor tracking and beyond.
With this paper we want to bring together browser vendors, advertising platforms, practitioners, and web users
to reflect on the tradeoffs between privacy and technological innovation,
and the impact that unethical practices may have on users in the real world.

\begin{acks}
We thank Heather O'Brien and Irene Lopatovska for providing early feedback,
as well as our anonymous reviewers.
\end{acks}

\appendix

\section{RNN code}

\autoref{fig:code} illustrates the deceptively ease of creating a fairly competent recurrent neural network
for two-class classification that takes as input a trajectory of $(x,y,t)$ coordinates (max. 100 timesteps)
and outputs the majority class probability.

\begin{figure}[!ht]
\small
\begin{kerascode}
model = Sequential()
model.add(Bidirectional(GRU(64), input_shape = (100,3)))
model.add(Dropout(0.25))
model.add(Dense(1, activation = "sigmoid"))
model.compile(loss = "binary_crossentropy", optimizer = Adam(0.0005))
\end{kerascode}
\vspace{-0.5em}
\caption{Our RNN implementation with the Keras library.}
\label{fig:code}
\end{figure}

\section{Resources}
\label{apx:resources}

We release a Chrome extension that implements our adversarial noise approach to distort the mouse cursor coordinates on the fly.
The extension can be enabled or disabled for whitelisted domains.
This way, the user can allow certain websites to track their mouse movements as needed;
e.g., as part of an auditing process of a banking website,
an e-commerce that do not request personal data but want to get a demographics overview of their visitors,
or simply a research study that pays the user for letting them to analyze their mouse cursor activity.
The extension can be downloaded at \url{https://github.com/luileito/mousefaker}.

Our dataset comprising mouse cursor movements and associated demographic variables
is available at \url{https://gitlab.com/iarapakis/the-attentive-cursor-dataset}.
Each user log includes the following information:
query, gender, age, browser viewport size (width and height),
and mouse cursor trajectory as a sequence of $(x,y,t)$ tuples.

\balance{}
\bibliographystyle{acmart}
\bibliography{refs}

\end{document}